\documentclass[pra, superscriptaddress, twocolumn, showpacs]{revtex4}
\usepackage{amssymb}
\usepackage{amsfonts}
\usepackage{epsfig}
\usepackage{amsmath}
\usepackage{graphicx}

\def\be{ \begin{equation} }
\def\ee{ \end{equation} }
\def\bea{ \begin{eqnarray} }
\def\eea{ \end{eqnarray} }
\def\bse{ \begin{subequations} }
\def\ese{ \end{subequations} }

\begin{document}

\title{Simulation of a quantum phase transition of polaritons with trapped ions }
\author{P. A. Ivanov}\email{pivanov@phys.uni-sofia.bg}
\affiliation{Institut f\"ur Quanteninformationsverarbeitung,
Universit\"at Ulm, Albert-Einstein-Allee 11, 89081 Ulm, Germany}
\affiliation{Department of Physics, Sofia University, James Bourchier 5 blvd, 1164 Sofia, Bulgaria}
\author{S. S. Ivanov}
\affiliation{Department of Physics, Sofia University, James Bourchier 5 blvd, 1164 Sofia, Bulgaria}
\author{N. V. Vitanov}
\affiliation{Department of Physics, Sofia University, James Bourchier 5 blvd, 1164 Sofia, Bulgaria}
\affiliation{Institute of Solid State Physics, Bulgarian Academy of Sciences, Tsarigradsko chauss\'{e}e 72, 1784 Sofia, Bulgaria}
\author{A. Mering}
\affiliation{Fachbereich Physik, Technische Universit\"at Kaiserslautern, D-67663 Kaiserslautern, Germany}
\author{M. Fleischhauer}
\affiliation{Fachbereich Physik, Technische Universit\"at Kaiserslautern, D-67663 Kaiserslautern, Germany}
\author{K. Singer}
\affiliation{Institut f\"ur Quanteninformationsverarbeitung,
Universit\"at Ulm, Albert-Einstein-Allee 11, 89081 Ulm, Germany}

\begin{abstract}
We present a novel system for the simulation of quantum phase transitions of collective internal qubit and phononic states with a linear crystal of trapped ions. 
The laser-ion interaction creates an energy gap in the excitation spectrum, which induces an effective phonon-phonon repulsion and a Jaynes-Cummings-Hubbard interaction. 
This system shows features equivalent to phase transitions of polaritons in coupled cavity arrays.
Trapped ions allow for easy tunabilty of the hopping frequency by adjusting the axial trapping frequency, and the phonon-phonon repulsion via the laser detuning and intensity. 
We propose an experimental protocol to access all observables of the system, which allows one to obtain signatures of the quantum phase transitions even with a small number of ions.
\end{abstract}

\pacs{03.67.Ac, 03.67.Lx, 03.67.Bg, 42.50.Dv}
\date{\today}
\maketitle


Trapped ions are among the most promising physical systems for implementing quantum computation \cite{NC} and quantum simulation \cite{FEYNM}. 
Long coherence times and individual addressing allow for the experimental implementation of quantum gates and quantum computing protocols such as the Deutsch-Josza algorithm,
 teleportation, quantum error correction, quantum Fourier transformation and Grover search \cite{GATESALGO}. 
Quantum simulation could be performed in the future by large-scale quantum computation \cite{KIELPINSKI02}. 
With the currently available technology, tailored Hamiltonians can be modeled with trapped ions to simulate mesoscopic
Bose-Hubbard systems \cite{PORRAS04}, spin-boson systems \cite{PORRAS08} and spin systems \cite{SPIN}. 
Recently, a quantum magnet consisting of two spins has been successfully simulated experimentally with two trapped ions \cite{FRIEDENAUER08}.

\begin{figure}[tbp]
\includegraphics[width=0.50\textwidth,height=0.25\textwidth,angle=0]{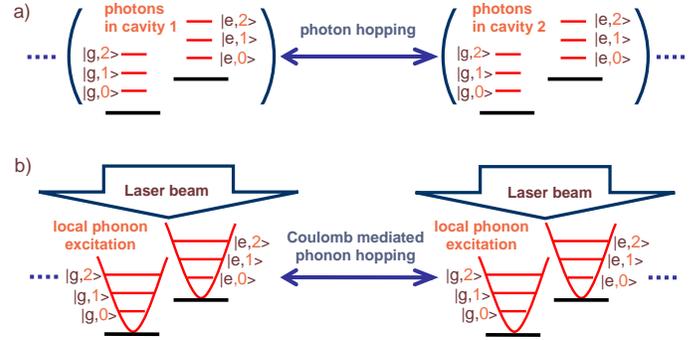} 
\caption{(Color online). a) Coupled cavities each containing photons and single two-state atoms. Intercavity hopping is provided by an optical fibre.
The strong coupling between the atoms and the photons leads to an effective photon-photon repulsion. b) All ions are simultaneously interacting with a traveling wave in the radial direction. 
The laser-ion interaction creates an effective on-site interaction between the local phonons. 
The phonon hopping appears due to the Coulomb interaction and can be adjusted by the mutual distance of the ions.}
\label{fig1}
\end{figure}

In this Letter, we propose a physical implementation of the Jaynes-Cummings-Hubbard (JCH) model using trapped ions. 
The JCH model was proposed in the context of an array of coupled cavities, each containing a single two-state atom and a photon \cite{G}. 
Such a system is described by the combination of two well-known physical models:
 the Hubbard model \cite{H,F}, which describes the interaction and hopping of bosons in an optical lattice, and the Jaynes-Cummings model,
 which describes the interaction of an atom with a quantum field \cite{WK}. 
The JCH model predicts a quantum phase transition of polaritons, which are collective photonic and atomic excitations. 
We shall show that the laser-driven ion chain in a linear Paul trap is described by a JCH Hamiltonian,
 wherein the ions and the phonons correspond, respectively, to the atoms and the photons in a coupled cavity array, Fig. \ref{fig1}.
As in \cite{PORRAS04}, the position-dependent energy and the non-local hopping frequency of the phonons is controlled by the trapping frequencies,
 while the effective on-site repulsion is provided by the interaction of the phonons with the internal states of the ions and can be adjusted by the parameters of an external laser field, namely the Rabi frequency and the detuning. 
This on-site interaction is analogous to the photon blockade (photon-photon repulsion), where the strong atom-cavity coupling prevents the entering of additional photons into the optical cavity \cite{ISWD}.
We shall show that many-body effects appear as a quantum phase transition
between a localized Mott insulator (MI) and delocalized superfluid state (SF) of the composite phononic and internal (qubit) states of the ions. 
Due to the collective nature of the excitations we distinguish between collective qubit and phononic SF and MI phases, and the pure phononic SF phase, similar to the effects predicted in \cite{IOK} for coupled cavity arrays.

Consider a chain of $N$ ions confined in a linear Paul trap along the $z$ axis with trap frequencies $\omega _{q}$ $\left( q=x,y,z\right) $,
 where the radial trap frequencies are much larger than the axial trap frequency ($\omega _{x,y}\gg \omega _{z}$), so that the ions are arranged
 in a linear configuration and occupy equilibrium positions $z_{i}^0$ along the $z$ axis. 
Making a Taylor expansion around the equilibrium position, and neglecting $x^{3},$ $y^{3},$ $zx^{2}$, $zy^{2}$ and higher order terms, the
Hamiltonian in the radial direction $x$ reads \cite{James} 
\begin{equation}
\hat{H}_x = \frac{1}{2M}\sum_{k=1}^{N}\hat{p}_k^2 + \frac{M\omega_x^2}{2} \sum_{k=1}^{N} \hat{x}_k^2
 - \frac{M\omega_z^2}{2} \sum_{\substack{ k,m=1 \\ k>m}}^{N} \frac{(\hat{x}_k-\hat{x}_m)^2}{|u_k-u_m|^3}.  \label{Hx}
\end{equation}
Here $\hat{p}_k$ is the momentum operator, $M$ is the ion mass, $\hat{x}_k$ is the position operator of the $k$th ion about its equilibrium position $z_k^0$,
 $u_k=z_k^0/l$, where $l=(e^2/4\pi \varepsilon_0M\omega_z^2)^{1/3}$ is the length scale, $e$ is the charge of the ion, and $\varepsilon_0$ is the permittivity in free space.
In the Hamiltonian \eqref{Hx} the motion in the radial direction is decoupled from the axial motion. 
In terms of the normal modes $\omega_p$, the Hamiltonian \eqref{Hx} reads $\hat{H}_x = \hbar \sum_{p=1}^{N}\omega_p (\hat{\widetilde{a}}{}_p^\dag \hat{\widetilde{a}}_p + \frac12)$.
Here $\hat{\widetilde{a}}{}_p^\dag $ and $\hat{\widetilde{a}}_p$ are the phonon creation and annihilation operators of the $p$th \textit{collective} phonon mode.
However, if $\hat{x}_k$ and $\hat{p}_k$ are written in terms of \textit{local} creation $\hat{a}_{k}^\dag$ and annihilation $\hat{a}_{k}$ phonon operators, so that
 $\hat{x}_k=\sqrt{\hbar/2M\omega_x}\,(\hat{a}_k^\dag +\hat{a}_k)$ and $\hat{p}_k=i\sqrt{\hbar M\omega_x/2}\,(\hat{a}_k^\dag -\hat{a}_k)$, the Hamiltonian \eqref{Hx} reads 
\begin{equation}
\hat{H}_x = \hbar \sum_{k=1}^N (\omega_x+\omega_k)\hat{a}_k^\dag \hat{a}_k + \hbar \sum_{\substack{ k,m=1 \\ k>m}}^N t_{km}(\hat{a}_k^\dag \hat{a}_m+\hat{a}_k\hat{a}_m^\dag),  \label{HxBH}
\end{equation}
where we have neglected higher-order (energy non-conserving) terms. 
The phonons are trapped with a position-dependent frequency 
\begin{equation}
\omega_k = -\frac{\alpha \omega_z}{2} \sum_{\substack{ s=1 \\ s\neq k}}^N\frac{1}{|u_k-u_s|^3},  \label{om_m}
\end{equation}
where $\alpha = \omega_z/\omega_x$, and they may hop between different ions, with non-local hopping strengths 
\begin{equation}
t_{km}=\frac{\alpha \omega_z}{2}\frac{1}{|u_k-u_m|^3}
\label{hopping}
\end{equation}
derived from the long-range Coulomb interaction \cite{PORRAS04}.

The collective and local creation and annihilation operators are connected by the Bogoliubov transformation, 
\begin{equation}
\hat{\widetilde{a}}{}_p^\dag = \sum_{k=1}^N b_k^{(p)}(\hat{a}_k^\dag \cosh\theta_p - \hat{a}_k \sinh\theta_p),  \label{bt}
\end{equation}
which preserves the commutation relation, $[\hat{a}_k,\hat{a}_m^\dag] = \delta_{km}$. 
Here $\theta_p=-\frac14\ln\gamma_p$, 
 with $\gamma_p=1+\alpha^2(1-\lambda_p)/2$ and $\lambda_p$ are the eigenvalues, with eigenvectors $\mathbf{b}^{(p)}$,
 of the matrix $A_{km}=\delta _{km}+2\sum_{s=1,s\neq k}(\delta_{km}-\delta_{sm}) /|u_k-u_s|^3$ \cite{James}. 
Using Eq.~\eqref{bt} one finds that the $p$th collective phonon state with zero phonons $|\widetilde0_p\rangle $ is a product of $N$
local squeezed states, $|\widetilde0_p\rangle =|\zeta_1^{(p)}\rangle \ldots |\zeta_N^{(p)}\rangle$, where 
\begin{equation}
|\zeta _{k}^{(p)}\rangle =\sum_{n_{k}=0}^{\infty }\sqrt{\frac{(2n_{k}-1)!!}{(2n_{k})!!}}\frac{(\tanh \theta _{p})^{n_{k}}}{\sqrt{\cosh \theta _{p}}}\,|2n_{k}\rangle .  \label{sq_state}
\end{equation}
Here $|2n_{k}\rangle $ ($k=1,2\ldots N$) is the local Fock state with $2n_k$ phonons. 
For the center-of-mass phonon mode we have $p=1$ and $\cosh\theta_p=1$;
 hence the collective ground state $|\widetilde0_1\rangle$ is a product of local ground states, $|\widetilde0_1\rangle=|0_1\rangle \ldots |0_N\rangle$. 
For a sufficiently small number of ions, we have $\cosh\theta_p\approx 1$ and $|\zeta_k^{(p)}\rangle \approx |0_k\rangle$. 
Since the lowest-energy collective vibrational mode in the radial direction is the highest mode $p=N$, we find that the superfluid ground state of the Hamiltonian \eqref{HxBH} is
\begin{equation}
|\Psi_{\text{SF}}\rangle = \frac{1}{\sqrt{N!}}\left( \sum_{k=1}^N b_k^{(N)}\hat{a}_k^\dag \right) ^{N}|0_1\rangle |0_2\rangle \ldots |0_N\rangle .  \label{SF}
\end{equation}
Here we have assumed the commensurate case where the number of ions is equal to the number of phonons. 
We find that the ratio between the average number of local phonons in the ground state is given by the square of the oscillation amplitudes:
 $\langle\hat{n}_k\rangle / \langle\hat{n}_m\rangle = (b_{k}^{(N)}/b_{m}^{(N)})^{2}$, where $\hat{n}_k = \hat{a}_k^\dag \hat{a}_k$ ($k=1,2,\ldots ,N$) is the local phonon number operator.

We shall show that the laser-ion interaction induces an effective repulsion between the local phonons. 
This interaction provides the phase transition from phononic SF state to composite SF and MI phases of the joint phononic and qubit excitations. 
Consider ion qubits with a transition frequency $\omega_0$, which interact along the radial direction with a common traveling-wave laser light addressing the whole ion chain with frequency $\omega_{\text{L}}$. 
The Hamiltonian of the system after the optical rotating-wave approximation is given by \cite{WML} 
\begin{equation}
\hat{H} = \hat{H}_x + \hbar\Omega\left[ \sum_{k=1}^{N}\hat{\sigma}_k^+\text{e}^{\text{i}\eta (\hat{a}_k^\dag + \hat{a}_k) - \text{i}\delta t} + \text{h.c.}\right] .  \label{H}
\end{equation}
Here $\hat{\sigma}_k^{+}=|e_k\rangle \langle g_k|$ and $\hat{\sigma}_k^-=|g_k\rangle \langle e_k|$ are the spin flip operators,
 $|e_k\rangle $ and $|g_k\rangle $ are the qubit states of the $k$th ion, $\Omega $ is the real-valued Rabi frequency,
 $\delta =\omega _{\text{L}}-\omega _0$ is the laser detuning, and $\eta =|\mathbf{k}|x_0$ is the Lamb-Dicke parameter,
 with $\mathbf{k}$ the laser wave vector, and $x_0=\sqrt{\hbar /2M\omega_x }$ is the spread of the ground-state wave function. 
The Hamiltonian, after transforming into the interaction picture by the unitary transformation $\hat{U} =\text{e}^{\text{i}\hat{H}_0t/\hbar}$,
 with $\hat{H}_0=-\hbar\omega_x \sum_{k=1}^{N}\hat{a}_k^\dag\hat{a}_k + \hbar \Delta \sum_{k=1}^{N}|e_k\rangle \langle e_k|$,
 in the Lamb-Dicke limit and after the vibrational rotating-wave approximation, reads 
\begin{eqnarray}
\hat{H}_I &=& \hbar \sum_{k=1}^{N}\omega _k\hat{a}_k^\dag\hat{a}_k+\hbar \Delta \sum_{k=1}^{N}|e_k\rangle \langle e_k|  \notag \\
&&+\hbar g\sum_{k=1}^{N}(\hat{\sigma}_k^{+}\hat{a}_k+\hat{\sigma}_k^-\hat{a}_k^\dag)  \notag \\
&&+\hbar \sum_{\substack{ k,m=1 \\ k>m}}^{N}t_{km}(\hat{a}_k^\dag\hat{a}_m+\hat{a}_k\hat{a}_m^\dag) - \mu\hat{N},  \label{HI}
\end{eqnarray}
where $\hat{H}_I = \hat{U}^\dag\hat{H}\hat{U} - \text{i}\hbar \hat{U}^\dag\partial_t\hat{U}$. 
We assume that the laser is tuned near the red motional sideband $\delta =-\omega_x -\Delta $,
 with a small detuning $\Delta$ ($\Delta \ll \omega_x$). 
The coupling between the internal qubit and local phonon states is $g=\eta \Omega $. The Hamiltonian \eqref{HI} is valid when $t_{km},g\ll \omega_x $,
 which ensures that higher terms, which violate the conservation of the total number of excitations, can be neglected. 
The first three terms in Eq.~\eqref{HI} describe the Jaynes-Cummings model. 
The first two terms correspond to the energies of the local phonons and the ions, while the third term describes the laser-ion interaction. 
The fourth term in Eq.~\eqref{HI} describes the non-local hopping of phonons between different ions and allows the comparison to Hubbard systems. 
The last term $-\mu\hat{N}$ is the usual chemical potential.


\begin{figure}[tbp]
\includegraphics[width=0.40\textwidth,height=0.30\textwidth,angle=0]{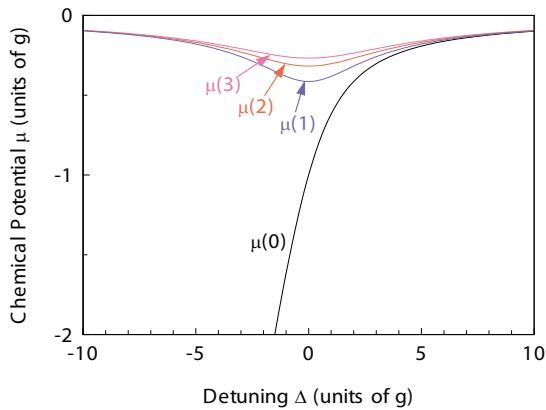}
\caption{(Color online) 
Chemical potential $\mu$ as a function of the laser detuning $\Delta$. 
Each curve corresponds to $\mu (n)=E_g(n+1)-E_g(n)$, where $E_g$ is the ground energy of the single ion Hamiltonian \eqref{HI} for zero hopping. 
Here $n$ is the number of excitations per ion. 
The Mott lobes are defined as $\delta \mu _k = \mu (k+1) - \mu (k),$ $k=0,1,2\ldots $}
\label{fig2}
\end{figure}

The Hamiltonian \eqref{HI} commutes with the total excitation operator $\hat{N} = \sum_{k=1}^{N} \hat{N}_k$, hence the total number of excitations is conserved. 
Here $\hat{N}_k=\hat{a}_k^\dag\hat{a}_k+|e_k\rangle \langle e_k|$ is the number operator of the total qubit and phononic excitations at the $k$th site. 
The effective on-site interaction is provided by the interaction of phonons and qubit states at each site. 
The strength of the on-site interaction depends on the external parameters, such as the Rabi frequency $\Omega $ and the laser detuning $\Delta $. 
This interaction creates an energy gap, which prevents the absorption of additional phonons by each ion. In Fig. \ref{fig2} we plot the chemical potential $\mu$,
 which counts the energy to add a extra excitation into the system, as a function of the laser detuning $\Delta $ for zero hopping. The Mott lobes $\delta \mu _{k}>0$ for
 sufficiently large number of phonons tend to zero, which show that the on-site interaction is vanishing \cite{G}.

We describe the quantum phase transition between the MI and SF states by the variance $\mathcal{D}N_k=\sqrt{\langle \hat{N}_k^2\rangle - \langle \hat{N}_k\rangle^2}$
 of the number operator $\hat{N}_k$ with respect to the ground state of the Hamiltonian \eqref{HI} for fixed number of excitations \cite{IOK}. 
If the on-site interaction between the phonons dominates the hopping, the ground state wave function is a product of local qubit and phononic states for each site with a fixed number of excitations. 
Hence in the MI state, the variance $\mathcal{D}N_k$ for any $k$ vanishes. 
When the hopping term dominates the on-site interaction, then the ground state consists of a superposition of qubit and phononic states with delocalized excitations over the entire chain. 
In this state the variance $\mathcal{D} N_k$ at each site (i.e. each ion) is non-zero.
 
\begin{figure}[tp]
\includegraphics[width=70mm,angle=0]{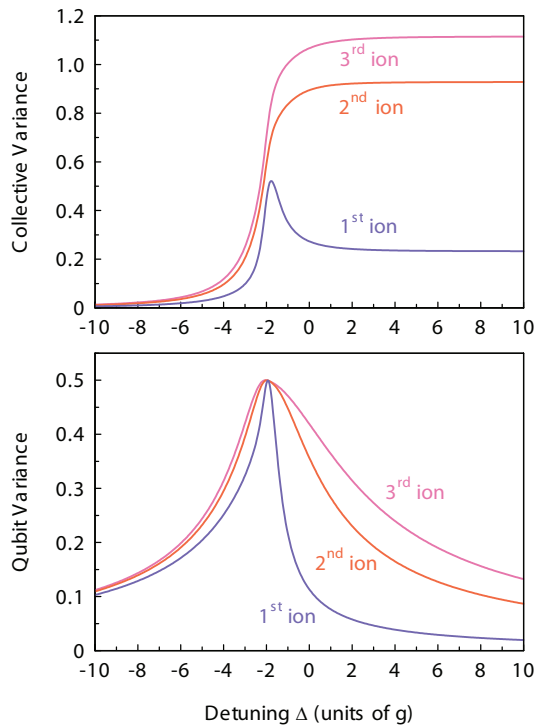}
\caption{(Color online) Total (qubit+phonon) variance $\mathcal{D}N_k$ (top) and the qubit variances $\mathcal{D}N_{a,k}$ ($k=1,2,3$) (bottom)
 for a chain of five ions with five excitations as a function of the laser detuning $\Delta $ for fixed hopping $t=0.3g$. 
Negative values of $\Delta$ correspond to blue detuning with respect to the red-sideband transition.}
\label{fig3}
\end{figure}

Figure \ref{fig3} (top) shows the variance $\mathcal{D}N_k$ ($k=1,2,3$) for a chain of five ions with five collective excitations
 versus the laser detuning $\Delta $ for fixed hopping frequency $t=\alpha \omega _{z}/2$ calculated by an exact diagonalization of the Hamiltonian \eqref{HI}. 
Due to the symmetry of the trap with respect to the center it is not necessary to plot the phase diagrams for ions $4$ and $5$.
For sufficiently large negative detuning $\Delta $, there exists an energy gap, which prevents the absorption of additional phonons. 
Hence, the system is in the MI phase, where the qubit and phononic excitation are localized.
When the detuning $\Delta $ increases, the energy gap decreases and the system makes a phase transition to the SF phase. 
The phase transition is stronger for the ions near the center of the trap due to stronger Coulomb interaction and increased hopping strengths, and weaker at the ends of the ion chain. 
The comparison between the variance at the different sites demonstrates two characteristic features. 
Firstly, there is a range of detunings where the qubit and phononic excitations at ion 1 (end of the chain) are predominantly in a MI phase, whereas the other ions are in the SF phase. 
Secondly, there is a broad range of $\Delta $ along which the joint excitations at all ions are in the SF phase.

Figure \ref{fig3} (bottom) shows the variance of the qubit excitations $\mathcal{D}N_{a,k}$ with $\hat{N}_{a,k} = |e_k\rangle \langle e_k| $, ($k=1,2,3$) \cite{IOK}.
This allows us to distinguish the following phases: in the region of large negative detuning $\Delta $ the collective and the qubit variances are small, indicating that the system is in the qubit MI phase. 
Increasing the detuning, the collective variance stays small but the qubit variance increases, which shows that the system is indeed in a collective MI phase.
Approaching $\Delta =0$ the system makes a phase transitions into the collective qubit and phononic SF phase as now both collective and qubit variance are large. 
Finally, for sufficiently large positive detuning the qubit variance decreases but the collective variance stays large, which shows that the system is in the phononic superfluid phase.

The experiment is started by initializing the ion chain with $N$ phonons in the lowest energy radial mode. To avoid off resonant excitation of unwanted radial modes, $\alpha$ could be increased temporarily. Then the coupling laser is switched on. The experimental proof for the phase transition can be carried out by local measurements repeated for each ion which should be performed faster than the hopping time. 
In the case of $^{40}$Ca$^{+}$ ions the qubit states could be represented by the ground state S$_{1/2}$ and the metastable state D$_{5/2}$. 
The laser, which creates the phonon-phonon repulsion, would be detuned to the red sideband of the quadrupole qubit transition between the two states. 
Then the readout could be performed by scattering photons on the dipole transition S$_{1/2}\rightarrow $P$_{1/2}$. 
This would lead to momentum recoil and changes of the phononic excitation, 
but to circumvent this we have to perform a measurement of the qubit states, which does not affect the phononic state as in the following steps.
1) Make a random guess for the qubit excitation. 
2) If the guess was the S$_{1/2}$ ground state, then swap the population S$_{1/2}\Leftrightarrow $D$_{5/2}$ by a carrier $\pi $ pulse leaving the phononic excitation unchanged. 
3) Now expose the ion to laser light on the dipole transition. 
4) If the ion scatters light the guess was wrong and we have to discard the measurement and restart, 
otherwise the initial guess was right and we transfer the qubit excitation back to the S$_{1/2}$ ground state by another carrier $\pi $ pulse,
 then drive Rabi oscillations on the red sideband by perpendicular Raman light beams with the difference momentum vector in the radial direction. 
6) The phononic population can now be extracted by a Fourier analysis of the Rabi oscillations.

In conclusion, we have proposed a novel implementation of the JCH model by trapped ions simulating polaritonic phase transitions in coupled cavity arrays. 
The system shows a Mott insulator to superfluid phase transition of the collective qubit and phononic excitation even with a small number of ions. 
The features can be easily measured by local laser adressing. 
Compared to atoms in optical cavities, our implementation is easier to manipulate, as all parameters can be tuned by changing the trap frequency, laser detuning and intensity.
Additionally, the system can be extended by adding impurities of ions with different mass to the ion crystals, which allows for 
simpler addressing of the radial phonon modes and a separation of coexistent phases. 

This work has been supported by the European Commission projects EMALI and FASTQUAST, the Bulgarian NSF grants VU-F-205/06, 
VU-I-301/07, D002-90/08, and the excellence programme of the Landesstiftung Baden-W\"{u}rttemberg. The authors
thank F. Schmidt-Kaler for useful discussions.


\end{document}